# PHYSICS OF THE OBSERVABLE. MECHANICS.

Adrián Faigón*

Device Physics Laboratory - Departamento de Física - Facultad de Ingeniería-
Universidad de Buenos Aires

Mechanics can be founded in a principle stating the uncertainty in the position of an observable particle $\delta q$ as a function of its motion relative to the observer, expressed in a trajectory representation. From this principle, $p.\delta q = \text{const.}$, being $p$ the q-conjugated momentum, mechanical laws are derived and the meaning of the Lagrangian and Hamiltonian functions are discussed. The connection between the presented principle and Hamilton's Least Action Principle is examined.

For a particle hidden from direct observation, the position uncertainty is determined by the enclosing boundaries, and is, thus, disengaged from its momentum. Heat, as a non-mechanical magnitude, stem from this fact, and thermodynamical magnitudes have direct expression in the presented formalism.

It is finally shown that in terms of Information Theory, mechanical laws have simple interpretation. Kinetic and potential energies are expressions of the information on momentum and position respectively, and the law of conservation of energy expresses the absence of information exchange in mechanical interactions.

----------

*A. Faigón is researcher at the CONICET (Consejo Nacional de Investigaciones Científicas y Técnológicas). email: afaigon@fi.uba.ar

# I. INTRODUCTION

Mechanics is the basis of the Physics building. In spite of their formal equivalence, each formulation of classical mechanics, Newton, Euler-Lagrange, Hamilton, Jacobi, contributes to clarify the fundamental concepts on which physics relies, and mediates the connection with other branches of Physics.[1]

Whichever formalism is used, Mechanics has to be supplemented with statistics to support Thermodynamics, thus bridging the unknown mechanical trajectories with observable thermodynamics results. [2]. Quantum Mechanics shares this feature with Thermodynamics, since both deal with the uncertainties inherent in any description of observables. Classical Mechanics, on the other hand is said to correspond to an infinitely precise description, or to be the limit of Quantum Mechanics for Planck's constant, h, tending to zero.

This contribution is an attempt to show that the above is not necessary, and that Classical Mechanics can be formulated in terms of the uncertainty associated with the description of observables. In addition to provide a new scope on mechanical laws, this formalism could aid in smoothing conceptual bridges between the main branches of Physics.

# II. THEORY

## A. CONSTANCY OF THE MOMENTUM – SPATIAL UNCERTAINTY PRODUCT.

The uncertainty in the spatial coordinate of an observable particle, $\delta q$, is related to a magnitude p, called the q-conjugated momentum, which characterizes its motion relative to the observer, through the constancy of their product. Representing this product by the function f of both variables, the above statement is expressed by

$$f(p, \delta q) \equiv |p| \cdot \delta q = const > 0. \qquad (1)$$

The meaning of δq is that the observer –the system with which the particle is in interaction—does not recognize two particle positions as distinct unless they are apart by δq or a greater distance. In this way, with regard to particle motion description, space appears segmented in δq's, which are, in general, position dependent.

Eq. (1) states the Principle of Constancy of the function f (PCF) for observable coordinates [3]

## B. DYNAMICS. THE TRAJECTORY DESCRIPTION.

The temporal evolution, motion dynamics, is obtained introducing time via the rate of change of f. From (1),

$$\dot{f} = |\dot{p}| \cdot \delta q + |p| \cdot (\dot{\delta q}) = 0 \quad . \qquad (2)$$

This result becomes the classical description, with the use of the trajectory representation, suggested by the motion of big bodies, as follows:

The trajectory representation consists of assigning a position q(t) to each time instant t. Thus, space and time are related to each other along the particle motion by the magnitude particle velocity,

$$\dot{q} \equiv \frac{dq}{dt}. \qquad (3)$$

In this representation, and in accordance with the constraints imposed by PCF, the motion occurs within a sheaf of indistinguishable trajectories bundled by the extremes q(t) and (q+δq)(t) as depicted in Fig.1. In this case, $(\dot{\delta q})$ in (2), satisfies

$$(\dot{\delta q}) = \delta \dot{q}, \qquad (4)$$

where the r.h.s. represents the difference in the particle velocity between the closest distinguishable trajectories.

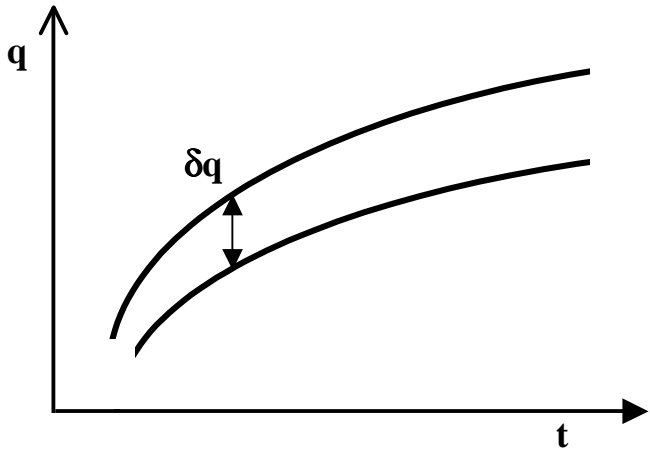

Fig. 1. The limits q(t) and (q+ δq)(t) of the sheaf of undistinguishable trajectories according to the PCF, eq. 1.

Substitution of (4) into the first equality of (2), yields

$$\dot{f} = |\dot{p}|.\delta q + |p|.\delta \dot{q} \equiv \delta £(q,\dot{q}). \qquad (5)$$

Last member denotes the difference between the values of a function of coordinates and velocities, $£(q, \dot{q})$, for the closest distinguishable trajectories at fixed time,

$$\delta £(q,\dot{q}) = \frac{\partial £}{\partial q} \cdot \delta q + \frac{\partial £}{\partial \dot{q}} \delta \dot{q},$$

if £ satisfies

$$\frac{\partial £}{\partial q} = |\dot{p}| \quad \text{and} \quad \frac{\partial £}{\partial \dot{q}} = |p| \quad ; \tag{6a, b}$$

in which case,

$$\frac{d}{dt}\left(\frac{\partial £}{\partial \dot{q}}\right) - \frac{\partial £}{\partial q} = 0 \quad . \tag{7}$$

Eqs. (6) shows that £ has the properties of the Lagrange function of classical mechanics, and eq. (7) is the corresponding equation of motion. Eq. (5) states that the broadest meaning of the Lagrange function lies in the fact that its variation between the extreme trajectories of the indiscernible sheaf equals the rate of change of the f product.

## C. PCF AND THE PRINCIPLE OF LEAST ACTION

The constancy of f may be written in integral form as

$$0 = f(t_2) - f(t_1) = \int_{t_1}^{t_2} \dot{f} \, dt = \int_{t_1}^{t_2} \delta £ \, dt = \delta \int_{t_1}^{t_2} £ \, dt = \delta A \Big|_{t_1}^{t_2} , \tag{8}$$

where A is the Hamiltonian action defined by

$$A \equiv \int £ \, dt \quad . \tag{9}$$

The result (8), formally identical to Hamilton's Principle of Least Action (PLA), was obtained as a consequence of the Principle of constancy of f (PCF), $\dot{f} = 0$, and from

(4) –i.e. with adoption of the trajectory representation—which leads to the identification (5) of δ£ with f.

The PLA in Hamilton's formulation states: The trajectory q(t) described by a certain degree of freedom of a system, given that q(t1) is q1 and q(t2) is q2, minimizes – extremalizes- the action defined in (9). I.e. the variation of A from trajectory q(t) to a neighboring q(t)+ δq(t), given

$$\delta q(t_1) = \delta q(t_2) = 0 \quad , \tag{10a}$$

must be zero.

The PCF states: Let δq be the uncertainty in the value of the position coordinate q, i.e. the minimum change in position which may be perceived by an observer (an interacting body); and p, the momentum; motion occurs in such a way that the product f=|p|.δq remains constant.

In the PLA, δq is a virtual and arbitrary variation of the trajectory; and the equations of motion stem from the restriction (10a) that nulls the term

$$p.\delta q \big|_{t1}^{t2} \tag{10b}$$

in the calculation of δA by parts.

In PCF, δq is a physically meaningful variable of the system (particle-interacting world), and the term (10b) in the calculus of δA vanishes with the constraint p.δq constant along the motion, which is less restrictive than (10a).

Thus, PLA may be interpreted as a particular case of PCF, requiring that at t1 and t2, i.e. any given instant, the product p.δq = 0, or, equivalently, that the trajectory be known with infinite precision. As it was shown, this latest condition, though sufficient, is not necessary. It suffices to require p.δq = constant, which is more adequate for a description of the observable, even in the absence of actual observations.

### D. THE HAMILTONIAN DESCRIPTION.

In terms of minimum distinguishable changes along the trajectory, the time derivative of any given magnitude y, will be evaluated by

$$\dot{y} = \frac{|\dot{q}|}{\delta q} \cdot \delta^- y \quad , \tag{11}$$

where $\delta^- y$ is the change in y while the particle moves from q to q+δq. Thus, the first equality in (2) may be written

$$\dot{f} = |\dot{q}| \cdot \delta^- p + \frac{|p||\dot{q}|}{\delta q} \cdot \delta^- \delta q \quad . \tag{12}$$

In the trajectory representation, and provided δq depends only on q, eq. (12) becomes

$$\dot{f} = |\dot{q}| \cdot \delta^- p + \frac{|p||\dot{q}|}{\delta q} \cdot \frac{\partial \delta q}{\partial q} \cdot \delta^- q \equiv \delta^- H(q,p) \quad , \tag{13}$$

which is the change in the value of the function H(q,p) while the particle passes from q to q+δq in the course of its motion, with H satisfying, thus,

$$\frac{\partial H}{\partial |p|} = |\dot{q}| \quad \text{and} \quad \frac{\partial H}{\partial q} = \frac{|p \cdot \dot{q}|}{\delta q} \cdot \frac{\partial \delta q}{\partial q} \quad . \tag{14 a,b}$$

The Hamiltonian function H, is related to the time evolution of f --eq.(13)—similarly to £ (eq. 5). In both cases, $\dot{f}$ equals the change of the corresponding function over the uncertainty interval δq. However, whereas the change in £ is between the closest indistinguishable trajectories at fixed time, the change in H is along the trajectory in the course of time and motion. To emphasize this distinction, $\delta^-$ is called change, whereas δ is referred to as a difference or a variation. To summarize:

$$\delta^- H = \delta £ = \dot{f} \qquad (15)$$

### E. THE MECHANICAL LAWS.

The mechanical laws are obtained from the general expressions of δ£ and $\delta^-$H for the case $\dot{f}=0$, and assigning the sign of $\dot{q}$ to p. In this case, eqs. (5), (6), and (12) to (14) are valid without the symbol of modulus. The requirement $\dot{f}=0$ in eq. (13) yields conservation of the Hamiltonian throughout motion.

Replacing the first term of (13) by, its original expression, $p \cdot \delta \dot{q}$, and using

$$\delta^- q = \delta q \qquad , \qquad (16)$$

results in

$$\dot{p} = -\frac{p \cdot \dot{q}}{\delta q} \cdot \frac{\partial \delta q}{\partial q} \qquad , \qquad (17)$$

which is Newton's second law, whereby the r.h.s. stands for the force (generalized force in case q is not a cartesian coordinate) acting on the body [4,5].

The two terms which add to $\delta^- H$ in (13) are termed the change in the kinetic energy $\delta^- E$, and potential energy $\delta^- V$, respectively; and their sum is the change in mechanical energy. Hence, conservation of mechanical energy is equivalent to $\dot{f} = 0$.

Substitution of eq. (17) in eq. (13), yield Hamilton's equations:

$$\dot{q} = \frac{\partial H}{\partial p} \quad \text{and} \quad \dot{p} = -\frac{\partial H}{\partial q} \quad , \tag{18 a,b}$$

which, unlike (14), express $\dot{f} = 0$ already included in (17). Similarly, if the second member of (13) defines $\delta^- H$ in general, for the mechanical case,

$$0 = \delta^- H = \dot{q}.\delta^- p - \dot{p}.\delta^- q \quad . \tag{19}$$

### F. THE MECHANICAL LAGRANGIAN

Comparing equation (19) with the $\delta £$ defined in eq. (5), it is apparent that formally, in the mechanical case, d£ and dH are related by

$$d£ (q,\dot{q}) = d(p\dot{q}) - dH(q,p) \quad , \tag{20}$$

which allows writing $\delta £$ -using (17)- as

$$\delta £ = p.\delta \dot{q} - \frac{p\dot{q}}{\delta q}.\frac{\partial \delta q}{\partial q}.\delta q \quad . \tag{21}$$

Whenever $p \cdot d\dot{q} = \dot{q} \cdot dp$ as is the case in classical mechanics, eq. (21) is the well-known Lagrangian form,

$$\delta £ = \delta E - \delta V \qquad (22)$$

In the relativistic case, in which momentum and velocity are not mutually proportional magnitudes, the kinetic term of the Lagrangian is obtained from the first term of eq. (21), consistent with the definition of £ in eq. (5). Using the relativistic momentum $p = m_o \dot{x} / \sqrt{1 - \dot{x}^2/c^2}$, it yields the known expression

$$\delta £ = \delta m_o c^2 \sqrt{1 - \dot{x}^2/c^2} - \delta V \qquad (23)$$

for the relativistic Lagrangian.

## G. A GATE TO THERMODYNAMICS.

If the particle does not interact freely with the rest of the world, but, instead, through the walls of a box which encloses it; the uncertainty interval δq will be determined by the box linear dimensions, Δq, and the f-product ceases to be constant since p and Δq may vary independently. The constant in eq. (1) is, in this case, the minimum value of f because Δq can not be less than the δq corresponding to the free particle [6].

As a consequence of unconstraining f, the more general expression for the rate of change of momentum in the trajectory representation is, from the first equality in eq. (2) and eq. (11),

$$\dot{p} = -\frac{p\dot{q}}{\delta q} \cdot \frac{\partial \delta q}{\partial q} + \frac{\dot{f}}{\delta q} \quad . \tag{24}$$

The first term is the general expression for the mechanical force. Eq. (24) indicates the limit of mechanics whenever causes appear for the change of p, other than the change in the localization of the particle δq, along the trajectory. These causes are expressed in $\dot{f} \neq 0$, which requires that the particle be hidden from direct observation.

Eq. (24) may be reinterpreted in energetic terms, if multiplied by δ-q, and using (11) and (16), to obtain

$$\dot{q}.\delta^- p = -\frac{p\dot{q}}{\delta q} \cdot \frac{\partial \delta q}{\partial q} .\delta^- q + \dot{f} \quad . \tag{25}$$

The l.h.s. is the change in internal energy, U, i.e. the energy associated with particles without direct interaction with the surroundings. Each term in the r.h.s. have particular meaning when the other is zero. The first is minus the work W done by the body on its surroundings, and the second is called the heat Q absorbed by the body. Formally,

$$\delta^- Q = \dot{q}.\delta^- p\big|_{\Delta q=cte} = \dot{f}\big|_{\Delta q=cte} = \frac{\dot{q}}{\Delta q}.\delta^- f\big|_{\Delta q=cte} \quad , \tag{26a}$$

and

$$\delta^- W = -\dot{q}.\delta^- p\big|_{f=cte} = \frac{p\dot{q}}{\delta q}.\delta^- \delta q\big|_{f=cte} \quad , \tag{26b}$$

and the decomposition

$$\dot{q}.\delta^- p = \dot{q}.\delta^- p\big|_{\Delta q=cte} + \dot{q}.\delta^- p\big|_{f=cte}$$

expresses the first law of thermodynamics, which states that heat is a work equivalent for energy exchanges,

$$\delta^- U = \delta^- Q - \delta^- W \quad .$$

Finally, it can be shown that, formally at least, entropy has a place in this formulation. Rewriting eq. (13)

$$\delta^- H = p\dot{q}.\frac{\delta^- f}{f} \quad , \tag{27}$$

and using the apparent equivalences

$$p\dot{q} \equiv kT \quad \text{and} \quad \delta^- \ln f \equiv \frac{\delta^- S}{k} \tag{28},(29)$$

where k is Boltzman's constant, T temperature, and S entropy; eq. (27) becomes

$$\delta^- H = T.\delta^- S \equiv \delta^- Q_r \quad , \tag{30}$$

identifiying H with what in thermodynamics is called reversible heat, Qr. Eqs. (26) to (30) define the main thermodynamic magnitudes.

### H. ACTION AND ENTROPY.

The integral expressing the variation of action between two instants equals the change of f in this interval, as shown in eq. (8). Thus, the least change in time for the variation of action is

$$\delta^-\delta A \equiv \delta A\Big|_{to}^{t(q+\delta q)} - \delta A\Big|_{to}^{t(q)} = \delta A\Big|_{t(q)}^{t(q+\delta q)} = \delta^- f \qquad , \qquad (31)$$

or, using (29),

$$\frac{\delta^-\delta A}{f} = \frac{\delta^- S}{k} \quad , \qquad (32)$$

which expresses the relationship between action and entropy [7,8].

Dividing eq. (32) by $\delta t \equiv t(q+\delta q) - t(q)$, yields

$$\delta\pounds = \frac{f}{k}.\dot{S} \quad , \text{ and, from (15), } \quad \delta^- H = \frac{f}{k}.\dot{S} \quad , \qquad (33 \text{ a,b})$$

which relate the variation in the Lagrangian, or the least change in the Hamiltonian with the rate of entropic change.

## I. DO THE EQUATIONS OF MOTION DERIVE FROM PLA? RETURNING TO THE MEANING OF PLA.

Two expressions were used for $\pounds$, both in the trajectory representation, namely, Eq. (5)

$$\dot{f} = |\dot{p}|.\delta q + |p|.\delta\dot{q} \equiv \delta\pounds(q,\dot{q}).$$

which, by construction, equals $\dot{f}$; and eq. (21), corresponding to the mechanical Lagrangian,

$$\delta\pounds^M = p.\delta\dot{q} - \frac{p\dot{q}}{\delta q}.\frac{\partial\delta q}{\partial q}.\delta q$$

which is equivalent to the former for $\dot{f}=0$. In case $\dot{f} \neq 0$, $\delta £^M$ no longer stands for $\dot{f}$, although it continues to be identically zero as evidenced by the application of (11) to $\delta q$,

$$(\dot{\delta q}) = \frac{|\dot{q}|}{\delta q} \delta^- \delta q = \frac{|\dot{q}|}{\delta q} \cdot \frac{\partial \delta q}{\partial q} \cdot \delta q,$$

and the use of (4) to replace $\dot{\delta q}$ in latter expression of $\delta £^M$.

If the Lagrangian $£^M$, e.g. Lagrange's original, $£^M = E-V$, has variation $\delta £^M = 0$, how can the equations of motion be deduced from PLA, which is, thus, trivially satisfied? The answer is as follows: In the conventional calculation of $\delta A$, one arrives to [1]

$$\delta A = p.\delta q \Big|_{t1}^{t2} - \int \left( \frac{d}{dt} \frac{\partial £}{\partial \dot{q}} - \frac{\partial £}{\partial q} \right) \delta q.dt$$

which, according to the above, is identically zero if $£$ is $£^M$. It is, thus, the cancellation of the first term, justified by the use of fixed extremes in the variational calculation, what yields the equation of motion. This justification is, as mentioned in section C, a concealed way, a way which does not complies with physics as the science of the observable, to express $\dot{f}=0$. To summarize, table I exhibits the relationships between $£$, $\dot{f}$, PLA and the equations of motion.

|  | $£$ from eq.(5) | $£^M$ from eq.(21) |
|---|---|---|
| always | $\delta £ = \dot{f}$ | $\delta £^M = 0$ |
| always | d/dt ∂£/∂q' - ∂£/∂q = 0 | $\delta A = 0$ |
| Mechanics: $\dot{f} = 0$ <==> | $\delta £ = 0$ | Equations of motion |

Table I. Validity of some important relationships for the Lagrangian defined in eq. 5, and the Mechanical Lagrangian of eq. 21.

## J. MECHANICS AND INFORMATION

Information theory entered Physics through the concept of entropy [9-11]. It does not permeate into Mechanics because in the very definition of information lies the change of some interval of uncertainty, a non-existent entity in the classic formulation which assumes infinite precision in the knowledge of every magnitude. It is, thus, natural, in the present formulation, which relies, on the contrary, on the indetermination of the fundamental quantities –eq. (1)-, to attempt an interpretation of Mechanics in terms of information.

Consistency is achieved by choosing |p| as the uncertainty interval, $\Delta p$, for the knowledge of p. Thus f is an uncertainty product [12], and the information, I, corresponding to the mechanical state of the particle, defined by its position and momentum, is the sum of information on position and momentum, i.e.:

$$dI \equiv dI_q + dI_p = -(d\ln \delta q + d\ln |p|) = -d\ln f, \qquad (34)$$

consistent with the relationship between information and entropy, and eq. (29) relating f to entropy.

Consequently, PCF, eq. (1), means that, in mechanical interactions, the mechanical information of each observable coordinate is conserved –i.e. the information that the interacting surroundings have about it-. Even more, as the value of f is minimum for mechanical interactions, the amount of information each part has about other is the maximum available –though not infinite as presumed in classical formulation-.

The absence of information exchange between the interacting parts of a mechanical system, or

$$dI=0 \qquad (35)$$

suggests the name Principle of Zero Information Exchange (PZIE) for eq. (1) in this framework.

In informational terms, mechanical laws have very simple forms. It can be verified, for example, that kinetic and potential energies, first and second terms of the second member in (13), represent information on momentum and position respectively, multiplied by kT; and the conservation of mechanical energy is, thus, a direct expression of PZIE. The limit of validity of Classical Mechanics is -footnote [5]-

$$\left|\delta^- I_q\right| = \left|\frac{\delta^- V}{kT}\right| \ll 1 \quad \text{or} \quad \left|\delta^- I_p\right| = \left|\frac{\delta^- E}{kT}\right| \ll 1, \qquad (36\ a,b)$$

similar to that setting the boundary between equilibrium and non-equilibrium thermodynamics.

17## III. CONCLUSIONS

It has been shown that Mechanical laws can be derived from the constancy of an action product, f, without the assumption of an infinitely precise description.

The Lagrangian and Hamiltonian functions acquire simple meanings in terms of the rate of change of the f-product, and the Hamilton Principle of Least Action results in a particular case of the law of constancy of f.

Thermodynamics appears as a simple extension of the formalism, allowing changes of the f-product in time. Thermodynamical magnitudes are, in this way, simply related with mechanical magnitudes, in particular entropy with action.

The formalism can be interpreted in terms of information theory, resulting the law of constancy of the function f, or the law of conservation of energy, equivalent to state that mechanical interactions occur without mechanical information exchange.

12. In the context of quantum mechanics, $\delta q$ is half De Broglie's wave length, and the constant in eq. (1) is half Planck constant, $h/2$). Interpreting f as an uncertainty product, classical mechanics is in the limit of the Heisenberg uncertainty principle, when the inequality turns to equality; without necessarily constraining h to tend to zero